# Flight in the Jovian Stratosphere – Engine Concept and Flight Altitude Determination

Nedislav S. Veselinov,[1]  Martin N. Karanikolov, [1] Vladislav V. Shihskin[1], Dimitar M. Mladenov[2]
*Sofia University "St. Kliment Ohridski", Sofia, Bulgaria*

**An effective method for detailed observation of the Solar System planets is the use of vehicles that can perform flight in their atmospheres, with the most promising of them being Flyers (aircraft for other planets atmospheres). Besides the advantage of probing the atmosphere directly, they have the ability to fly on selected direction and altitude, making them suitable for collecting information over large areas. Equipping the Flyer with nuclear propulsion will allow it to conduct flight for months without the need of combustible fuel or oxidizer to be carried on board. Among the planets of the Solar System and their satellites, Jupiter is a viable target for exploration, since it features thick atmosphere suitable for aerodynamic flight, there is no solid surface that can be contaminated after end of the mission, and the atmospheric data for designing a Flyer is readily available. This paper proposes a mathematical model for evaluating the thrust, the lift and the maximum allowable mass for horizontal steady flight as functions of the altitude and different heat chamber temperatures.**

## I. Introduction

FROM the beginning of the space era, over thirty probes have been sent into the atmosphere or landed on the surface of the other planets in the Solar System. However, most of them lacked the capability for sustained flight. The only vehicles that successfully reached and conducted continuous flight in the atmosphere of a planet other than the Earth were the Vega-1 and Vega-2 balloons in 1985. They flew over twenty thousand kilometers into the atmosphere of Venus and collected valuable scientific information [1, 2]. NASA's Perseverance rover or the planned Dragonfly mission [3] will utilize rotary aircraft. However, these will rely on electric motors and will have very limited range and speed.

---

[1] PhD Student, Faculty of Physics, Department of Radio Physics and Electronics.
[2] Dr., Associated professor, Faculty of Physics, Department of Theoretical Physics.

Among remote observation, satellites remain the most widespread mean of exploring extraterrestrial atmospheres. However, the satellites fly high above the dense atmosphere. The most efficient way to probe and assess important physical parameters like pressure and temperature distribution, and altitude gradients, is to perform flight through different layers of the atmosphere. This can be achieved by flying the satellite into the thick lower layers towards the end of its mission or by designing an atmospheric entry spacecraft to collect data during its descent. Two notable examples for such spacecraft are the ESA's Huygens lander on Titan and NASA's Galileo spacecraft, which uploaded valuable atmospheric data from Jupiter for almost an hour during its descent [4,5].

Continuous, powered flight can be conducted by a Flyer that relies on classical airplane principles. To produce sufficient lift, relatively dense atmosphere and / or high airspeed are required. The engine will need to produce enough thrust to enable high-speed flight. The flight duration will obviously be limited to the amount of propellant carried on board or the mission will need to rely on external energy source, like solar power. However, the usage of electric engines supplied by solar panels is meaningful only near bodies close to the Sun. For instance, the energy reaching the Jovian atmosphere consists of about 5% of the solar irradiation intensity at Earth's orbit [6]. Furthermore, only fraction of the solar energy reaches the lower layers, due to atmospheric scattering and other effects. For this reason, electric propulsion relying on solar panels is an unfeasible choice for high-speed flight on bodies with dense atmosphere or beyond the orbit of Mars.

An alternative approach would be to utilize nuclear heat to produce thrust. The nuclear fuel has extremely high energy density that allows for months, if not years, of sustainable flight before the fuel is depleted. Unlike chemical combustion, the nuclear reaction does not rely on oxygen to produce heat. This enables flight in anaerobic atmospheres and without the need of carrying oxidizer. The engine can be designed as a ramjet, which relies on supersonic gas compression instead of turbo compressor to produce thrust, which has a number of advantages: it has few moving parts, which minimizes the risk of mechanical failure, and it is light. The latter is of paramount importance, given the capabilities of the launch vehicles and the cost to deliver every kilogram into orbit of other planets[3]. Such design is called Nuclear-Powered Ramjet Engine (NPRE).

A nuclear heat engine has been tested on Earth within the US Military Project Pluto and has shown very promising results. The engine achieved 156,000 N thrust during tests [7]. The project envisaged the creation of the SLAM

---

[3] The mission cost of the Galileo mission was $1,5 billion [8,9]. Given the mass of the atmospheric probe and the satellite of about 3.000 kg, means that the per kg mission cost was approx. $500.000

(Supersonic Low Altitude Missile) nuclear-powered supersonic missile, capable of performing long-duration flight on complex trajectories. The project was closed in 1964, because the military favored the intercontinental ballistic missile approach. However, the results show that the technical challenges are manageable and NPRE is a viable flight propulsion option.

Research on nuclear-powered planetary flight was conducted in Ref [8]. The work suggests that a flight could be performed with a 3-tonne Flyer at subsonic speed. The proposed Flyer featured an engine with a turbo compressor, nuclear heat chamber and a turbine. It relied on a classical turbo-jet principle to produce thrust but relied on a nuclear heat chamber instead of chemical combustion. Considering the gas composition of Jupiter, a very high rotational speed and multiple compressor stages will be required to achieve sufficient compression. Such compressor will need to withstand higher mechanical loads and will have higher mass, compared to a turbo machine for similar conditions on Earth. This will increase the total mission mass, cost and the risk of mechanical failure. For these reasons, ramjet engines relying on supersonic compression are more practical compared to subsonic turbo designs, especially for flight in hydrogen-rich atmospheres with high local speed of sound.

Another in-depth research on nuclear-powered flight was conducted in Ref. [9, 10]. In this research, an atmospheric variation of MITEE (*MIniature reacTor EnignE*) nuclear rocket engine was proposed. Some physical and geometric parameters of the engine were calculated, without conducting detailed design or evaluating possible flight altitudes. The proposed design features small payload and very high engine temperatures (1500 K) which would require a cooling solution and very powerful reactor.

A NPRE-based Flyer has its technical limitations. It requires dense atmosphere for sufficient thrust and lift to be generated. This makes the concept unsuitable for flight on Mars, for instance. In case of rocky bodies with dense atmospheres like Venus and Titan, there is the moral obstacle of using nuclear power for aerodynamic flight, since the Flyer will ultimately crash into the surface and contaminate to local ecosystem with radioactive material. These considerations make gas giants a viable option for such a mission. They feature thick atmospheres with no hard surface and are particularly interesting for exploration, due to the presence of weather and different atmospheric phenomena.

Jupiter has several advantages making it suitable for such a mission over the other gas giants in the Solar System: it is closer to Earth and easier to reach; its atmospheric and wind conditions are less aggressive, compared to Saturn, Uranus and Neptune; and its atmospheric composition has been studied by the Galileo probe, which facilitates the Flyer design.

During the initial stages of the research, the topic of developing a Flyer suitable of sustained flight in the atmospheres of gas giants and Jupiter were met with interest from the research society. This is due to the fact that, besides being an interesting topic, the studies in this field are limited, with very little research publicly available. By determining the thrust and lift requirements for horizontal steady flight as function of the altitude and heat chamber temperature, this work provides the first necessary tool for the preliminary design of the Flyer.

The derived equations enable the definition of the Flyer external configuration and its payload, provide important input for the planning of the scientific experiments and mission operations, and allow the determination of the carrier rocket requirements.

In this paper, a mathematical model for the determination of possible flight altitudes in the Jovian atmosphere is introduced. The required thrust for steady flight at different altitudes of an idealized ramjet engine is calculated as a function of the temperature in the heat chamber. The maximum possible Flyer mass is derived. The model provides the necessary input for the initial design process. After the initial design, the detailed construction of engines and Flyers can be carried out, which is subject of future work.

In the second part of this paper, the atmospheric conditions in the flight area are described, and the boundary conditions for the calculation are defined. In the third part, the calculation method is described. The fourth part includes the conclusion and offers an insight into possible future work.

## II. Gas Composition and Environment

An altitude or flight level on Jupiter can be defined as the elevation above the Jovian Sea Level JSL (the JSL is the isobaric surface of static pressure equal to the Earth's Mean Sea Level pressure or 1013,25 hPa). Most suitable for aerodynamic flight are the lower part of the stratosphere and the upper part of the troposphere up to 60 km above JSL. The pressure and density in this range match the atmospheric conditions on Earth at 20-30 km altitude, which is within the service ceiling of typical supersonic aircraft. However, compared to Earth, a Flyer will experience about 2,5 times higher gravitational acceleration, meaning it will either need to produce 2,5 higher lift or be 2,5 lighter to achieve sustainable flight (gravitational acceleration at JSL is $g_{Jup} = 24,79 \ m/s^2 = 2,53 \cdot g_{Earth}$) [5].

The absolute upper limit for the calculations was set at the altitude of 90 km above the JSL, because the density on that level ($\rho_{90km} = 0,0018 \ kg/m^3$) is equivalent to the air density on Earth at 50 km AMSL. 23,5 km was the chosen lower limit, because this is the lowest altitude for which the Galileo probe returned full set of data.

The gas in the range of interest consists of approximately 86,1 % Hydrogen and 13,6% Helium by mass fraction. The composition does not change considerably between 23,5 km and 90 km: the molecular weight and specific gas constant are constant, while the heat capacity ratio varies by ±2% from an average of 1,54. The speed of sound gradually decreases at lower altitudes, meaning the Flyer will produce more thrust at lower flight levels.

The haze layer upper limit is at around 50 km above JSL and there is ammonia cloud layer at up to around 30 km that contains ammonia ice particles. The lower edge of the ammonia cloud layer is at around 0 km. The flight must be conducted preferably above the cloud layers, to allow unobstructed visual observation [11].

### III. The Calculation Approach

The flight performance calculation is based on an idealized engine with 0,5 m² cross-section. The producible thrust and maximum allowable Flyer mass are calculated as a function of the desired altitude and for different heat chamber temperatures. A schematic of a NPRE is shown on Fig. 1. The operating principle of the engine relies on supersonic compression. During flight, the atmospheric gas enters the engine through the supersonic inlet 2, where it is slowed down through series of supersonic shocks. The gas continues to diffuse in the diffusor 3 before entering the heat chamber, where it is heated up by the reactor the design temperature 4. The hot, highly compressed gas passes the nozzle throat 5, and is subsequently accelerated to high supersonic speed, producing thrust.

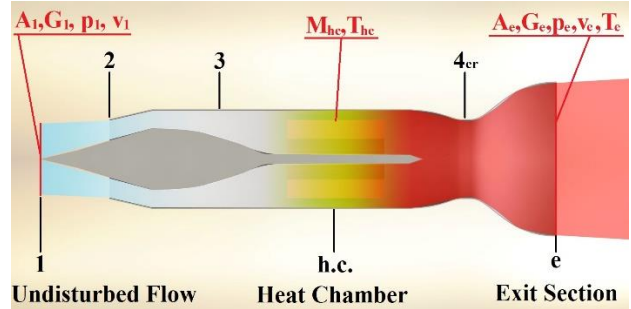

**Fig. 1 NPRE sections: 1 – Undisturbed flow, 2 – Gas inlet, 3 – Diffusor, h.c. – Heat chamber, 4cr – Critical nozzle section, e – Nozzle outlet (exit section). The notation for the relevant physical parameters for each section is also given**

The thrust R produced by the NPRE can be calculated from the following relation [12, 13]:

$$R = G_e v_e - G_1 v_1 + (p_e - p_1)A_e - X_\alpha, \qquad (1)$$

where $G_1, p_1, v_1$ and $G_e, p_e, v_e$ are atmospheric gas mass flow rate, pressure and velocity of the undisturbed flow (index "1") and at the nozzle outlet (index "e") respectively, $A_e$ is the nozzle outlet surface area and $X_\alpha$ is an additional drag term.

In an ideally designed inlet, the geometry is optimized in a way to allow the oblique shock waves originate from a common point on the front edge of the inlet. The ideal diffusion angles which lead to such shock structure can be determined experimentally in a wind tunnel or by performing Computational Fluid Dynamics (CFD) simulations. Under these conditions, the $X_\alpha$ term will be close to zero.

The second term $(p_e - p_1)A_e$ is the pressure thrust term. For a nozzle with finite size, the jet flow will not be ideally expanded, i.e. it will leave the engine at a higher pressure than the ambient one, producing thrust. In an ideal engine the gases leave the domain fully expanded. In this case the pressure thrust term is fully converted into momentum thrust and the engine efficiency is maximized.

The first $G_e v_e - G_1 v_1$ term is the momentum thrust term. This term can be maximized by introducing a nozzle with suitable expansion ratio to allow for full expansion of the gases leaving the engine.

For this calculation, an idealized highly efficient NPRE is assumed and both $X_\alpha$ and the pressure term are considered negligible. For a non-combusting engine in stable operation the gas mass flow rate leaving the engine equals the inlet flow rate, i.e. $G_e = G_1$. The thrust relation will then take the following form:

$$R = G_1(v_e - v_1) \tag{2}$$

The undisturbed flow speed is determined by the flight Mach number ($M_1$) and the speed of sound of the environment ($c$), which can be calculated from the Galileo probe data:

$$v_1 = cM_1 \tag{3}$$

High supersonic speed is required for good compression. At flight Mach number $M_1 = 3,0$ the free flow speed at an altitude of 60 km is 2469 m/s.

The mass flow rate depends on the engine inlet area $A_1$:

$$G_1 = A_1 \rho_1 v_1 \tag{4}$$

The required cross section area will depend on the shock wave structure and gas compression efficiency. For the current idealized estimate, $A_1$ is assumed to be 0,5 m². The density $\rho_1$ is available from the Galileo probe data.

The speed of the exhaust flow $v_e$ can be determined by the following expression:

$$v_e = M_e\sqrt{\gamma\, R_j T_e} \tag{5}$$

The exhaust flow speed needs to be higher than the flight speed. However, higher $M_e$ will require larger and heavier nozzle. The exhaust Mach number is assumed $M_e = 3{,}2$ which provides good balance between thrust and resulting mass. The heat capacity ratio $\gamma$ and specific gas constant $R_j$ are taken from Galileo data.

The exhaust temperature $T_e$ is a function of the heat chamber temperature $T_{hc}$ and the heat chamber to exhaust flow Mach number ratio, i.e. $T_e \sim f\left(M_{hc}/M_e\right)$. Since the NPRE is a breathing engine, the gas flows through the heat chamber at non-zero speed $M_{hc}$. The heat chamber Mach number will depend on the geometry and resulting shock structure and needs to be estimated. Obviously, for a ramjet, $M_{hc}$ will be between 0 and 1. In this analysis, $M_{hc}$ is assumed to be 0,5. The exhaust temperature can be calculated from the following relation [12]:

$$T_e = T_{hc}\frac{1+\frac{\gamma-1}{2}M_{hc}^2}{1+\frac{\gamma-1}{2}M_e^2} \tag{6}$$

With equations (1) to (6), the thrust can be calculated for different altitudes above JSL. Table 1 shows the results at altitudes between 30 km and 90 km and at four different heat chamber temperatures $T_{hc} = 600K; 900K; 1200K$ and $1500K$. The thrust as function of altitude is shown in Fig. 2.

**Table 1 Flight parameters and resulting thrust at different altitudes and for varying heat chamber temperatures**

| Altitude above JSL, km | Galileo probe data | | Engine mass flow rate, $G_1$, kg/s | Undisturbed flow speed, $v_1$, m/s | Exit temperature, $T_e$, K | Exit velocity, $v_e$, m/s | Thrust, $R$, N |
|---|---|---|---|---|---|---|---|
| | Pressure, Pa | Density, kg/m³ | | | | | |
| Heat chamber temperature – $T_{hc}=600K$ | | | | | | | |
| 90,4 | 1058 | 0,002 | 2,6 | 2774,9 | 178,6 | 3143,0 | 947,3 |
| 80,1 | 1632 | 0,003 | 4,2 | 2655,4 | 176,3 | 3133,9 | 1999,9 |
| 70,4 | 2587 | 0,005 | 6,9 | 2569,0 | 173,8 | 3123,1 | 3825,5 |
| 60,0 | 4374 | 0,010 | 12,2 | 2468,8 | 171,4 | 3113,7 | 7893,4 |
| 50,2 | 7620 | 0,019 | 22,3 | 2377,8 | 169,3 | 3104,5 | 16189,2 |
| 40,2 | 13450 | 0,033 | 39,4 | 2380,2 | 168,7 | 3101,7 | 28406,7 |
| 30,1 | 24150 | 0,061 | 71,3 | 2345,8 | 170,8 | 3111,9 | 54626,4 |
| Heat chamber temperature – $T_{hc}=900K$ | | | | | | | |
| 90,4 | 1058 | 0,002 | 2,6 | 2774,9 | 267,9 | 3849,3 | 2765,3 |
| 80,1 | 1632 | 0,003 | 4,2 | 2655,4 | 264,4 | 3838,2 | 4943,8 |
| 70,4 | 2587 | 0,005 | 6,9 | 2569,0 | 260,7 | 3825,1 | 8670,6 |
| 60,0 | 4374 | 0,010 | 12,2 | 2468,8 | 257,1 | 3813,5 | 16458,0 |
| 50,2 | 7620 | 0,019 | 22,3 | 2377,8 | 253,9 | 3802,2 | 31734,4 |
| 40,2 | 13450 | 0,033 | 39,4 | 2380,2 | 253,0 | 3798,8 | 55849,9 |
| 30,1 | 24150 | 0,061 | 71,3 | 2345,8 | 256,1 | 3811,3 | 104492,2 |

| | | Heat chamber temperature – $T_{hc}$=1200K | | | | | |
|---|---|---|---|---|---|---|---|
| 90,4 | 1058 | 0,002 | 2,6 | 2774,9 | 357,2 | 4444,8 | 4297,9 |
| 80,1 | 1632 | 0,003 | 4,2 | 2655,4 | 352,6 | 4432,0 | 7425,6 |
| 70,4 | 2587 | 0,005 | 6,9 | 2569,0 | 347,6 | 4416,8 | 12755,2 |
| 60,0 | 4374 | 0,010 | 12,2 | 2468,8 | 342,8 | 4403,4 | 23678,3 |
| 50,2 | 7620 | 0,019 | 22,3 | 2377,8 | 338,6 | 4390,4 | 44839,7 |
| 40,2 | 13450 | 0,033 | 39,4 | 2380,2 | 337,3 | 4386,5 | 78985,6 |
| 30,1 | 24150 | 0,061 | 71,3 | 2345,8 | 341,5 | 4400,9 | 146531,1 |
| | | Heat chamber temperature – $T_{hc}$=1500K | | | | | |
| 90,4 | 1058 | 0,002 | 2,6 | 2774,9 | 446,5 | 4969,4 | 5648,1 |
| 80,1 | 1632 | 0,003 | 4,2 | 2655,4 | 440,7 | 4955,2 | 9612,1 |
| 70,4 | 2587 | 0,005 | 6,9 | 2569,0 | 434,5 | 4938,1 | 16353,9 |
| 60,0 | 4374 | 0,010 | 12,2 | 2468,8 | 428,5 | 4923,2 | 30039,6 |
| 50,2 | 7620 | 0,019 | 22,3 | 2377,8 | 423,2 | 4908,6 | 56385,7 |
| 40,2 | 13450 | 0,033 | 39,4 | 2380,2 | 421,7 | 4904,3 | 99368,5 |
| 30,1 | 24150 | 0,061 | 71,3 | 2345,8 | 426,9 | 4920,4 | 183568,0 |

Since during steady flight the thrust equals the drag, the lift can be calculated from the thrust, if the lift-to-drag – L/D ratio is known. The maximum L/D ratio $k_{max}$ can be calculated from the Kuchemann's relationship [14]:

$$k_{max} = \left(\frac{L}{D}\right)_{max} = \frac{4(M+3)}{M} \qquad (7)$$

where $M$ is the flight Mach number. For $M = 3$, $k_{max} = 8$. However, this is an idealized empirical calculation valid on Earth. The exact L/D ratio will depend on many factors (Flyer geometry and configuration, angle of attack, etc.) and needs to be determined during detailed design by means of experiment or CFD simulations.

For this analysis, the L/D ratio will be assumed to be $k = 2,5$. This is considered a conservative estimate with the real value expected to be higher.

The lift is calculated as follows:

$$L = kR = 2,5R \qquad (R = D) \qquad (8)$$

The lift as function of altitude at different heat chamber temperatures is shown in Fig. 3

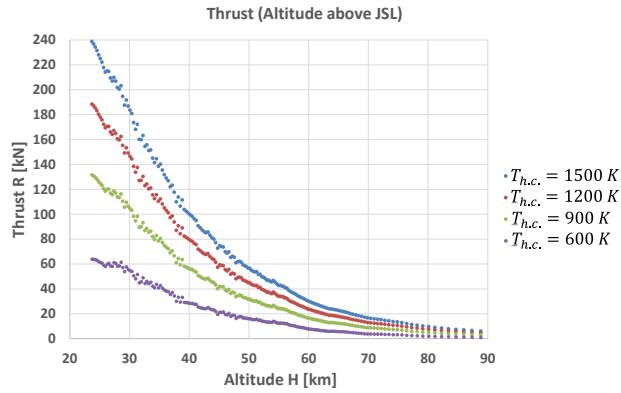

**Fig. 2 Resulting thrust as function of altitude for different heat chamber temperatures**

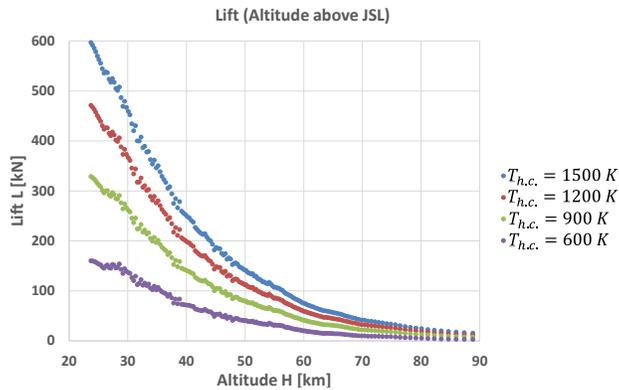

**Fig. 3 Resulting lift as function of altitude for different heat chamber temperatures**

In order to determine the maximum allowable Flyer mass, the weight needs to be calculated. The gravitational acceleration changes insignificantly within the altitude range of interest and is assumed constant at 23,2 m/s². The resulting Flyer masses for different temperatures $T_{h.c.}$ = 1500$K$; 1200$K$; 900$K$; 600$K$ and at different altitudes are shown in Fig. 4.

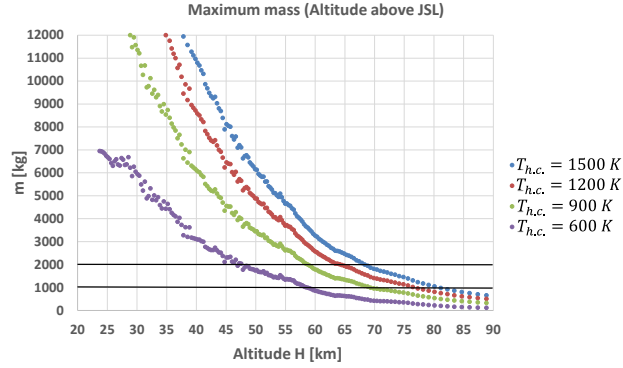

**Fig. 4** Maximum allowable mass for horizontal flight as function of altitude and for different heat chamber temperatures. The horizontal lines represent two hypothetical Flyers with 1000 kg and 2000 kg mass

The distributions show that heavier Flyers will require a higher heat chamber temperature for horizontal flight at given altitude. Table 2 shows a summary of the altitudes for steady flight for a 1000 kg and 2000 kg Flyer at different heat chamber temperatures.

**Table 2 Summary of the altitudes for steady flight for a 1000 kg and 2000 kg Flyer at different heat chamber temperatures**

| $T_{hc}$, K | Altitude of steady flight with Flyer mass of 1000 kg, km | Altitude of steady flight with Flyer mass of 2000 kg, km |
|---|---|---|
| 1500 | 81 | 68 |
| 1200 | 77 | 64 |
| 900  | 69 | 58 |
| 600  | 58 | 47 |

The flight altitudes are graphically shown on Fig. 5. It is evident that a Flyer with 1000 kg mass and reactor capable of producing 600 K in the heat chamber will be capable of flying at 58 km and above the haze layer.

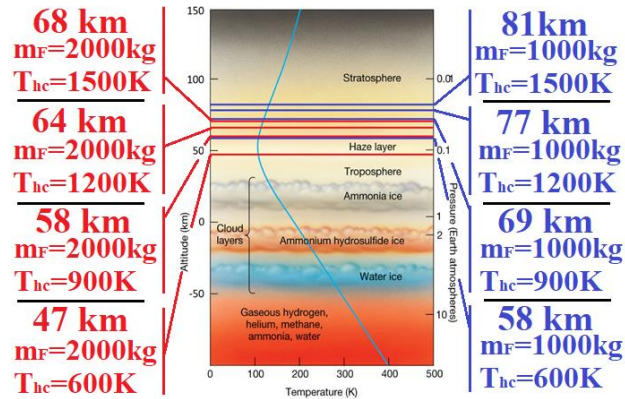

**Fig. 5  Flight altitudes for 1000 kg and 2000 kg Flyers and at different heat chamber temperatures. Background picture source https://www.uccs.edu/ ; Pearson Education, Inc.**

## IV. Conclusion and Future Work

The conducted calculation shows that flight with Nuclear-Powered Ramjet Engine (NPRE) is possible in the Jovian atmosphere. A NPRE can produce sufficient thrust for sustained horizontal flight at heat chamber temperatures as low as 600 K. The analysis shows that the lower layers of the Jovian stratosphere are sufficiently dense to allow for efficient engine operation.

The calculations are based on assumed compression efficiency, without calculating the supersonic diffusion explicitly. The performance is evaluated directly from the undisturbed flow parameters and depends on reference engine size to determine the mass flow. A real ramjet engine would require a smaller cross-section than the one assumed for this analysis, due to compressibility effects.

The cruise flight lift and maximum allowable mass are calculated out of assumed lift-to-drag – L/D ratio. The accurate L/D ratio will impact the operating altitude but can only be determined after more detailed design is conducted.

For these reasons, the results from this analysis serve as an approximate estimate. However, the determined parameters provide the required input for more detailed analysis. A more accurate assessment of the performance of the Flyer will require an explicit gas compression model of the engine. The model needs to allow calculation the mass flow, pressure losses and gas parameters throughout the engine, in order to determinate the thrust.

Another task is to determine the optimal engine inlet geometry for the desired flight altitude. This can be achieved by conducting CFD simulations. After the inlet performance is known, it will be possible to calculate the required reactor power for achieving the various heat chamber temperatures and resulting thrust levels.

After the thrust is defined, the flight performance can be determined more accurately. Work can proceed on the Flyer external configuration (outer body, aerodynamic surfaces, experimental payload configuration, etc.). It is essential to determine the L/D ratio accurately and for different flight conditions and altitudes as well as to assess different flight scenarios.

The presented methodology can be used for the development of engines and Flyers for atmospheric flight on the Solar System's other celestial bodies like Saturn, Uranus, Neptune, Venus and Titan. Apart from the scientific interest in the exploration of celestial bodies, there is also another emerging interest: the in-situ resource utilization. Concerning the presence of large quantities of powerful energy resources, such as He-3 on some of the planets, Flyer missions may prove to be crucial for further space exploration [15].

## Conflict of interest

The authors declare that they have no conflict of interest.

## Appendix

Some Abbreviations:

JSL – Jupiter Sea Level = 0 km – The level of pressure equal to 1 atm

NPRE – Nuclear-Powered Ramjet Engine

Flyer – An aircraft conducting flight in the atmosphere of other celestial body

CFD - Computational Fluid Dynamics

## Funding Sources

This work was partially supported by the Bulgarian Ministry of Education and Science under the National Research Programme "Young scientists and postdoctoral students" approved by DCM # 577/17.08.2018, by the Sofia University Research Fund under Grant No. 80-10-139/15.04.2019 and by the Sofia University Student Council Fund.

# References

[1] R.Z. Sagdeev, V.M. Linkin, V.V. Kerzhanovich, A.N. Lipatov, A.A. Shurupov, J.E. Blamont, D. Crisp, A.P. Ingersoll, L.S. Elson, R.A. Preston, C.E. Hildebrand, B. Ragent, A. Seiff, R.E. Young, G. Petit, L. Boloh, Y.N. Alexandrov, N.A. Armand, R.V. Bakitko, A.S. Selivanov, "Overview of VEGA Venus Balloon in Situ Meteorological Measurements", *Science* 231 (4744), 1986.

doi: 10.1126/science.231.4744.1411

[2] J. Blamont, R. Young, A. Seiff, B. Ragent, R. Sagdeev, V. Linkin, V. Kerzhanovich, A. Ingersoll, D. Crisp, L. Elson, R. Preston, G. Golitsyn, V. Ivanov, "Implications of the VEGA Balloon Results for Venus Atmospheric Dynamics", *Science* 231(4744), 1986.

doi: 10.1126/science.231.4744.1422

[3] Lorenz, R.D. & Turtle, E.P. & Barnes, J.W. & Trainer, M.G. & Adams, D.S. & Hibbard, Kenneth & Sheldon, C.Z. & Zacny, K. & Peplowski, P.N. & Lawrence, D.J. & Ravine, M.A. & McGee, T.G. & Sotzen, K.S. & MacKenzie, S.M. & Langelaan, Jack & Schmitz, Sven & Wolfarth, L.S. & Bedini, P.D.; "Dragonfly: A rotorcraft lander concept for scientific exploration at titan"; *Johns Hopkins APL Technical Digest (Applied Physics Laboratory)*; vol. 34; 2018; pp. 374-387.

[4] Johnson, T. V., Yeates, C. M., & Young, R.; "Space Science Reviews volume on Galileo Mission overview"; *Space Science Reviews*, vol. 60, no. 1-4, 1992, pp. 3-21.

doi: 10.1007/BF00216848

[5] A. Seiff, T.C.D. Knight, R.F. Beebe and L.F. Huber, GP-J-ASI-3-ENTRY-V1.0, *NASA Planetary Data System*, URL: https://pds-atmospheres.nmsu.edu/PDS/data/gp_0001/ [retrieved 07 March 2019]

[6] Li L, Jiang X, West RA, Gierasch PJ, Perez-Hoyos S, Sanchez-Lavega A, Fletcher LN, Fortney JJ, Knowles B, Porco CC, Baines KH, Fry PM, Mallama A, Achterberg RK, Simon AA, Nixon CA, Orton GS, Dyudina UA, Ewald SP, Schmude RW Jr. "Less absorbed solar energy and more internal heat for Jupiter", *Nature Communications,* vol 9(3709), 2018

doi: 10.1038/s41467-018-06107-2

[7] Frank E. Rom, *"Analysis of a Nuclear-Powered Ramjet Missile"*, Research memorandum, Lewis Flight Propulsion Laboratory, Cleveland, Ohio, (1954); URL: https://digital.library.unt.edu/ark:/67531/metadc60302/m2/1/high_res_d/19930088171.pdf [retrieved 15 May 2019]

[8] K. Miller, "Planetary flight", *Journal of Propulsion and Power* 11(5), 1063, 1995.

doi: 10.2514/3.23936

[9] J.R. Powell, J. Paniagua, G. Maise, H. Ludewig, M. Todosow, "Missions to the Outer Solar System and Beyond", *Acta Astronautica* 44(2), 159, 1999.

doi: 10.1016/S0094-5765(99)00043-0

[10] G. Maise, J. Powell, J. Paniagua, E. Kush, P. Sforza, H. Ludewig, T. Dowling, "Application of the MITEE Nuclear Ramjet for Ultra Long Range Flyer Missions in the Atmospheres of Jupiter and the Other Giant Planets" *54th International Astronautical Congress of the International Astronautical Federation, the International Academy of Astronautics, and the International Institute of Space Law*, Bremen, Germany, 2003.
doi: 10.2514/6.IAC-03-Q.4.09

[11] A. Ingersoll, T. Dowling, P. Gierasch, G. Orton, P. Read, A. Sánchez-Lavega, A. Showman, A. Simon, Amy, A. Vasavada. "Dynamics of Jupiter's Atmosphere", *Jupiter. The Planet, Satellites and Magnetosphere*, 2004, pp. 105-128.

[12] Loh, W.H.T. Jet, Rocket, *Nuclear, Ion and Electric Propulsion: Theory and Design*, 1967;
doi: 10.1007/978-3-642-46109-5

[13] M. Bondaryuk and S. Il'yashenko, *Ramjet Engines* (*Pryamotochnyye Vozdushno-Reaktivnyye Dvigateli*) Gosudarstvennoye Izdatel'stvo Oboronnoy Promyshlennosti, Moscow, 1958

[14] Antonella Ingenito, Stefano Gulli, Claudio Bruno; "Preliminary Sizing of Hypersonic Airbreathing Airliner"; *Transactions of the Japan Society for Aeronautical and Space Sciences*, Aerospace Technology Japan, vol. 8, 2010;
doi: 10.2322/tastj.8.Pa_19

[15] Palaszewski B. "Atmospheric Mining in the Outer Solar System: Outer Planet Resource Processing, Moon Base Propulsion, and Vehicle Design Issues", *AIAA Propulsion and Energy 2019 Forum*, Indianapolis, IN, 2019
doi: 10.2514/6.2019-4031